# High-performance ultrafast pulse compression in the visible spectral range for extreme nonlinear optics at kHz – MHz repetition rates

**SIYANG WANG[1], JIEYU YAN[1], SIRIUS SONG[1], ALEXANDER ATANASSOV[1], ZHIHAN WU[1], WILL BRUNNER[1], DIMITAR POPMINTCHEV[2], TENIO POPMINTCHEV[1,2,3]**

[1]*University of California, San Diego, La Jolla, CA 92093, USA*
[2]*Photonics Institute, TU Wien, Vienna A-1040, Austria*
[3]*tenio.popmintchev@physics.ucsd.edu*

**Abstract:** We demonstrate a remarkably effective single-stage compression technique for ultrafast pulses in the visible electromagnetic spectrum using second-harmonic pulses at 515 $nm$ derived from a 1030 $nm$ Yb-based femtosecond regenerative amplifier. By employing an advanced multi-plate scheme, we achieve more than fourfold compression from 180 $fs$ to 40 $fs$ with an extremely high spectral broadening efficiency of over 95%, and a temporal compression efficiency exceeding 75%. In addition, our method leverages a low nonlinearity medium to attain the shortest pulse durations for a single compressor while maintaining a superb spatial beam quality with 97% of the energy confined in the main lobe of the Arie disk. Moreover, our technique enhances the temporal pulse quality at 515 $nm$ without generating substantial femtosecond-to-picosecond pulse pedestals. The resulting intense visible laser pulses with excellent spatio-temporal parameters and high repetition rate of 100 $kHz$ to 1 $MHz$ open up new frontiers for extreme nonlinear optics and ultrabright EUV and X-ray high-harmonic generation using short VIS wavelength.

## 1. Introduction

Fully spatially and temporally coherent light with ultrashort wavelengths and pulse durations is essential for studies of ultrafast dynamics in atomic and molecular systems, advanced nanomaterials, plasmas, and bio systems [1-12]. One of the most promising techniques for producing such light is the process of high-order harmonic generation (HHG), which involves the upconversion of UV-VIS-IR pulses from femtosecond lasers or optical parametric amplifiers (OPAs) to extreme ultraviolet (EUV) or soft X-ray frequencies [1-4, 6, 8, 9, 13-16]. While in general, high harmonic generation typically requires high peak-power pulses with optimal $3 - 10$ laser cycles to reach high cutoff photon energy and record conversion efficiency greater than $10^{-3}$ - $10^{-7}$ using UV - to - mid-IR drivers [8, 9, 15, 16], simple post-compression techniques to reduce the laser pulses durations further, at shorter UV – VIS driving laser wavelengths, are in strong demand. Several methods have been shown to compress femtosecond pulses at near-infrared wavelengths to few-cycle durations, however, a practical scheme for spectral broadening and compression of UV – VIS pulses at high peak and high average power has not been reported yet [17, 18]. Moreover, most laser systems typically used for HHG generation operate at low repetition rates of up to a few kHz. At the same time, many EUV – X-ray imaging and spectroscopic applications would benefit from a high repetition rate high-flux tabletop source.

In this paper, we present a technique for generating high-power UV-VIS pulses with several cycle pulse duration optimal for EUV – soft X-ray applications with high photon energy and efficiency. Specifically, we demonstrate that a kHz-to-MHz Yb:KGW sub-picosecond amplifier at 1030 nm can be used for highly efficient high harmonic generation in the EUV region by spectrally broadening and compressing its second harmonic at 515 nm. The HHG process in gases driven by short-wavelength VIS lasers combines several advantages - very high single-atom efficiency due to low quantum diffusion of the rescattering electron, enhanced phase and group-delay matching due to high linear and nonlinear indices of refraction of atoms and ions, ultra-narrow linewidths of the harmonics and additional boost of the macroscopic efficiency due to broader temporal phase-matching window of 10s of laser cycles [9]. Moreover, the excellent spatial coherence and the extended soft X-ray cutoff with intrinsically-compressed near-transform limited attosecond pulses make this technique very attractive for high-resolution dynamic imaging and angle-resolved photoemission spectroscopies [7]. Finally, using enhanced UV-VIS laser beam parameters optimized for high harmonic generation in gas-filled capillaries could benefit from self-confinement of the driving pulses in both space and time. Here, we demonstrate a near spatio-temporal solitary propagation mode in periodic thin-plate media at VIS laser wavelengths and achieve spectral broadening via self-phase modulation (SPM) while maintaining an excellent spatial profile of nearly-identical intensity and similar pulse durations at each plate. This eliminates the substantial conical emission loss and enhances the efficiency of the pulse broadening geometry to above 93%, resulting in compression from $180\,fs$ to $40\,fs$ FWHM at $515\,nm$, or 23 laser cycles, with 42 µJ post-compression pulse energy when using a prism-pair compressor. This pulse duration is optimal for efficient, fully phase-matched, high-order harmonic generation using shorter-wavelength UV – VIS drivers due to the extended phase matching window in this regime. The pulse compression in a prism pair is designed to minimize the higher order dispersion of the spectrally broadened pulses using an analytical Lah-Laguerre optimization method and can be further improved to optimize both the transmission and compression of the pulses by using chirped mirrors with custom designed $2^{nd} - 4^{th}$ dispersion orders based on advanced dispersion calculations.

## 2. Spectral Broadening in Low Nonlinearity Multi-Plate Geometry

Self-phase modulation, arises from the laser-induced third-order Kerr nonlinearity. It modifies the spectral content of a pulse by changing the refractive index of the material $n(t) \cong n_0 + n_2 I(t)$ based on the intensity of the laser $I(t)$, and the nonlinear refractive index $n_2$. Here, we choose Calcium Fluoride ($CaF_2$) as a nonlinear medium because of its large bandgap and relatively low linear and nonlinear refractive indices [19]. The nonlinear accumulated phase, often referred to as B-integral, can be evaluated as:

$$B(t) = \frac{\omega_0}{c} \int_0^\ell n_2\, I(z,t)dz \approx \frac{2\pi}{\lambda} \ell n_2 I(t) \qquad (1)$$

, where $\omega_0$ is the central pulse frequency, $c$ is the speed of light, and $\ell$ is the thickness of the media. This nonlinear phase results in a frequency shift in the laser spectrum: $\Delta\omega(t) = -\frac{\partial}{\partial t}B(t)$. For a Gaussian pulse, this results to a maximum spectral broadening $\Delta\lambda$ due to self-phase modulation of approximately $\Delta\lambda = \Delta\lambda_0\sqrt{1+(0.88B)^2}$ [20]. Correspondingly, as a rule of thumb, a compression to a shorter pulse duration $\tau$ on the order of $\tau = \tau_0/\sqrt{1+(0.88B)^2}$ can be easily achieved with a second order phase compensation of $GDD \cong 0.88\tau_0^2 B/4\ln(2)$. The intensity slope of the laser field predetermines the maximum frequency change. In general, tightly focused beams with intensities lower than the damage critical intensity of the media can produce considerable spectral broadening with some high order dispersion being hard to be compensated. However, using spatial-temporal solitons at moderate intensities is more beneficial for self-phase modulation and pulse compression. Although there is an accumulation of nonlinear and linear phases due to the properties of the media, an external compressor can yield pulses close to the Fourier transform limit with a small amount of unbalanced higher dispersion orders. Nevertheless, the spectral broadening process can also cause temporal splitting of the laser pulse, or the media can significantly reshape the temporal and spatial profile of the laser, making it difficult to efficiently compress using standard chirp compensation techniques. In our design, we use thin plates of Calcium Fluoride cut in the [111] direction to minimize temporal splitting and substantial spatial deformation. For a broad range of laser wavelengths and relatively high dispersion orders, $CaF_2$ provides smaller dispersion compared to most commonly used materials, such as sapphire, fused silica, etc. Group delay dispersion (GDD) and third-order dispersion (TOD) are the most significant pulse-reshaping factors in this spectral range, and can be easily evaluated using [21-23]:

$$POD(n) = \frac{\partial^p}{\partial \omega^p} k(\omega) = (-1)^p \frac{1}{c} \left(\frac{\lambda}{2\pi c}\right)^{p-1} \sum_{m=0}^{p} \mathcal{B}(p,m) \lambda^m \frac{\partial^m}{\partial \lambda^m} n(\lambda) \qquad (2)$$

The GDD and TOD at central wavelength 515nm are relatively small: $48.619 \frac{fs^2}{mm}$ and $16.744 \frac{fs^3}{mm}$, respectively.

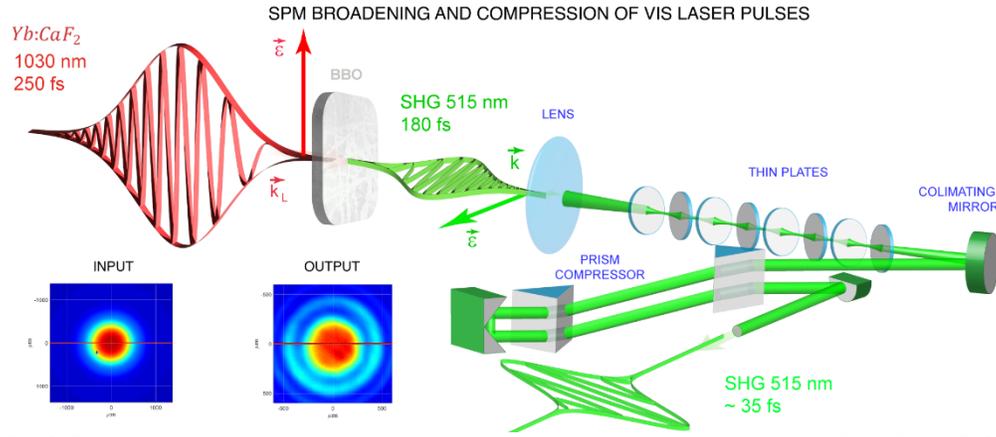

**Fig. 1. Schematic of high performance, single-stage, spectral broadening, and compression of ultrafast VIS 515 nm laser pulses using thin plates with low nonlinearity.** A high power $8\,W$, $250\,fs$, near-IR laser pulses from a KGW regenerative laser amplifier are upconverted with 70% efficiency to its second harmonic of $5.6\,W$, $180\,fs$, $515\,nm$ pulses in a BBO crystal. The VIS pulses are spectrally broadened in an array of thin $CaF_2$ plates with low nonlinearity in alternating Brewster angle geometry, with a high 93% efficiency. Then the pulses are compressed using a prism-pair compressor to sub-40 $fs$ with an excellent Gaussian beam profile with 78% efficiency. The overall efficiency is 75%. The inset shows the generated input SHG beam and the output SHG beam after the prism-pair compressor.

A single thin plate can provide a significant spectral broadening by tight focusing, however, accumulating a substantial nonlinear phase usually leads to a conical emission with an Arie ring pattern causing extensive energy losses. In addition, any formation of a single or multiple filaments inside the solid material leads to severe beam distortions or beam splitting. Alternatively, a larger beam waist can reduce the fast nonlinear phase accumulation per plate with a weaker self-refocusing. Thus, a considerable spectral broadening should preferably be realized using a soliton-like propagation in a set of thin plates. In our experiments, a 1030 nm $Yb{:}KGW$ laser amplifier produces $250\,fs$ laser pulses with 8 W average power and a tunable repetition rate of $100\,kHz$ - $1\,MHz$. These pulses are focused into a second-harmonic Type I BBO crystal with a high conversion efficiency of ~70% (see Fig. 1). The $515\,nm$ beam of $56\,\mu J$ energy per pulse at $100\,kHz$ and a shorter $180\,fs$ pulse duration, due to the nonlinear intensity dependence in upconversion, is then focused by a long focal length lens of $F = 1\,m$, ensuring a long Rayleigh range of interaction of more than 18 $cm$ [24]. While our 1030 nm amplifier has negligible femtosecond-to-picosecond pedestal in the time domain, the second harmonic generation process cleans any weak intensity temporal structure due to the nonlinear dependence on the intensity. Eight $CaF_2$ thin plates of 1 $mm$ thickness are used as a spectral broadening Kerr medium, equally spaced by $L_s = 50\ mm$, with the first plate placed right at the flat wavefront of the beam focus. The plates are aligned at alternating Brewster angles to minimize reflection loss and to compensate for any wavefront distortion. Additionally, the precise [111] $CaF_2$ crystal cut eliminates any polarization degradation and energy loss since $CaF_2$ is known to possess wavelength-dependent depolarization in white-light generation for other orientations.

One fascinating occurrence in self-confinement pulse propagation is the refocusing cycle of the laser. During laser self-confinement or filamentation, the beam rapidly shrinks down and expands, leading to a continuous series of divergence and refocusing that can persist over long distances. The formation of self-confinement generates plasma that transforms a narrowband laser pulse into a broadband pulse. An intriguing feature of this plasma is its ability to restrict the density of electrons, thereby averting optical breakdown. In a medium with Kerr nonlinearity, the focal length of self-focusing, to a second order approximation, scales with the square of the initial beam waist $w_0$, and is inversely proportional to the nonlinear refractive index of the medium $n_2$ and its length $\ell$, as well as the laser peak intensity $I$: $f_c \approx \frac{w_0^2}{4n_2 I \ell} = \frac{z_R}{2B}$. In experiments, where tight focusing is needed, it can be beneficial to set the distance between the plates to approximately $f_c-2f_c$, where $f_c$ could vary on each plate, and fine tune the laser intensity. The nonlinear accumulated phase in Eq. (1) sets a restriction on the intensity length product for desired nonlinear phase accumulation of $I\ell \left[\frac{W}{cm^2} mm\right] \cong 1.6 \cdot 10^{-7} \frac{B[rad] \cdot \lambda[nm]}{n_2[cm^2/W]}$, while the damage threshold for $CaF_2$ of near $P \sim 1 GPa$, sets a limit to the intensity of approximately $I_{dam} = Pc \cong 29.9 \frac{TW}{cm^2}$, where the speed of light is $c \cong 0.299 \frac{GW}{N}$. In our SPM design, the intensity on the first plate is $I \cong 601.2 \frac{GW}{cm^2}$ (peak power $P_p \cong 292.3 MW$) with a nonlinear phase $B = 1.43\ rad$, and a Kerr lens focal length $f_c \cong 65.9 mm$. The critical power for self-focusing of $CaF_2$ at the laser wavelength is $P_{cr} = \lambda^2 \frac{0.61^2 \pi}{8 n_2 n_0} \cong 1.3\ MW$. In our setup, we require a plate separation $L_s$ much smaller than the Rayleigh range to ensure a better control of the propagation in a near plane-wave geometry: $z_R = \frac{\pi \omega_0^2}{\lambda} > L_s$. Our theoretical design based on constraints of the nonlinear phase per plate suggests compression to a transform-limited pulse of 35 $fs$ when using the minimum of four thin plates assuming a dispersion compensation of the second order only with an expected total value of $GDD \cong 2300\ fs^3$.

### 3. High-Performance VIS Laser Pulse Compression with Enhanced Spatial and Temporal Quality

A large number of nonlinear phenomena related to third-order susceptibility affect the spatio-temporal propagation of intense laser light in media, making it challenging to optimize spectral broadening. In this work, we focus on experimentally maximizing the SPM and self-focusing at minimum high order phase distortions or wavefront distortions, and enhancing the spatio-temporal laser properties for applications in extreme nonlinear optics. To maintain a spatial soliton-like propagation, each plate must contribute to the SPM-induced spectral broadening and provide similar self-focusing. As the laser spot size on each plate, monitored using $2f - 2f$ imaging, reaches $350 \pm 10$ μm and remains unchanged, and the pulse duration after each plate measured using an autocorrelator and an scanning FROG apparatus (a self-diffraction and second harmonic FROG) stays approximately $200 \pm 10\ fs$, the VIS laser pulse enters a spatio-temporal soliton-like mode that supports spectral broadening without splitting. This contrasts with many SPM regimes where strong asymmetric spectral modulations are observed experimentally. To minimize the spatio-temporal distortions, the B-integral for each plate must be nearly identical and low, such that each plate contributes to the spectral broadening while minimizing any distortions. Note that the gas environment also plays a role despite its nonlinearity being substantially smaller than the solid materials. A soliton mode of propagation in multiple plates in vacuum, in combination with controlled pressure and gas species, is expected to fine-tune the dispersion of the SPM continuously, in a non-discrete way. In addition, atomic or molecular gases with different polarization properties or different nonlinearity can be used to adjust the spectral broadening, the blue and red shift of the VIS spectrum, to enhance the spatial-spectral-temporal beam quality with a real time feedback from the extreme nonlinear optics experiments.

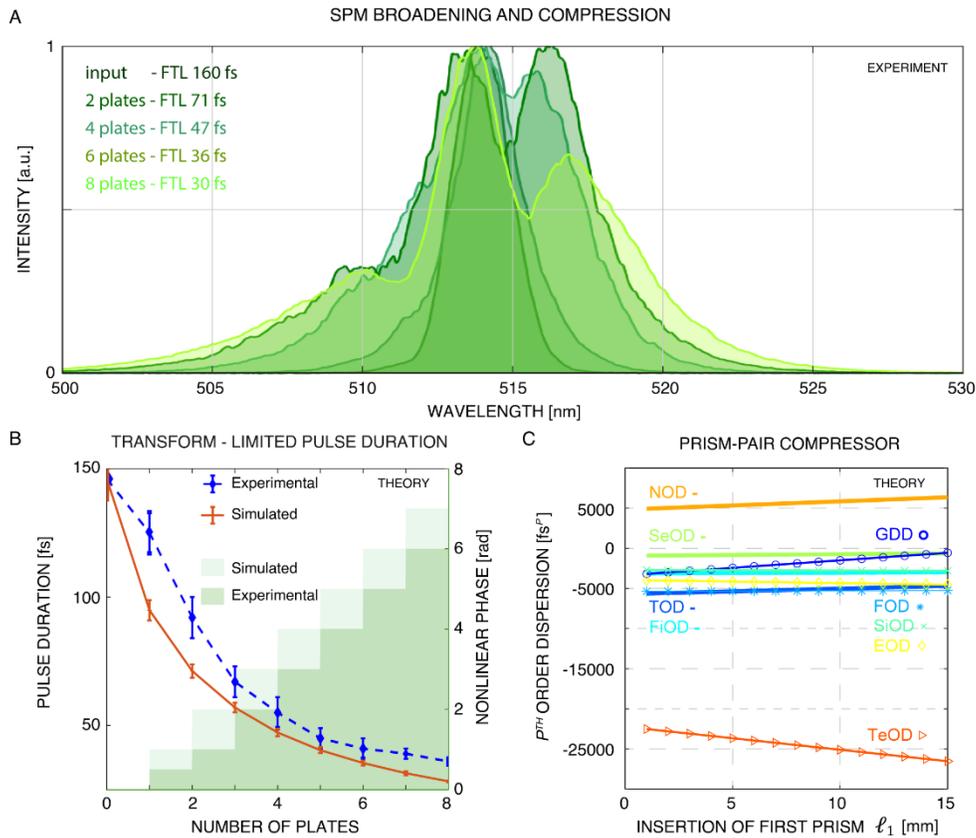

**Fig. 2. Experimental self-phase-modulation induced broadening of the spectrum of VIS femtosecond pulses in thin plates with fine spectral modulations.** A. Spectral broadening after each $CaF_2$ plate in the wavelength and frequency domain. The broadening is symmetric at low B integral and becomes slight asymmetric for higher cumulative nonlinear phase originates due to the initial dispersion of the second harmonic pulse. B. Rapid decrease of the Fourier Transform Limited (FTL) pulse duration after each plate with asymptotic decrease of the slope after $6^{th}$ – $8^{th}$ plates. The broadest VIS spectrum which supports a 35 $fs$ transform-limited pulse duration (20 laser cycles) is compressed by a prism-pair compressor to sub-40 fs (23 laser cycles). The calculated and experimental nonlinear phase and corresponding B-integral after each plate are shown in green. C. Design of the prism-pair compressor using Lah-Laguerre formalism for $GDD \sim 2000 - 3000\ fs^2$ and relatively low values of the high orders of dispersion (up to the tenth order) based on the numerically evaluated phase. Dispersion orders: GDD – $2^{nd}$, TOD – $3^{rd}$, FOD – $4^{th}$, FiOD – $5^{th}$, SiOD – 6h, SeOD – $7^{th}$, EOD – $8^{th}$, NOD – $9^{th}$, TeOD – $10^{th}$.

FROG trace measurements can retrieve the spectral phase information as a direct experimental characterization of the amount of phase compensation needed at the different dispersion orders. Here, we start with an analytical calculation to estimate the dispersion required to cancel the second order while minimizing the third-order dispersion and potentially higher orders [21-23]. The Group Delay Dispersion (GDD) and next immediate higher-orders of dispersion (TOD, FOD, etc.) of the prism-pair compressor are easily evaluated [21-23], Fig. 2C:

$$POD(n) = \frac{\partial^p}{\partial \omega^p} \varphi(\omega) = (-1)^p \frac{1}{c}\left(\frac{\lambda}{2\pi c}\right)^{p-1} \sum_{m=0}^{p} \mathcal{B}(p,m) \lambda^m \frac{\partial^m}{\partial \lambda^m} OP(\lambda) \quad (3)$$

where $OP$ is the optical path of the prism-pair compressor, evaluated here for a pulse with finite spectral content [25]. A fused silica prism pair compressor with a 70 cm tip-to-tip distance and with the amount of insertion material as shown in Fig. 2C, can compensate for the low orders of dispersion (predominantly GDD with partial minimization of the immediate higher orders) and yield a sub-40 $fs$ pulse with superb spatial and temporal quality (Fig. 3).

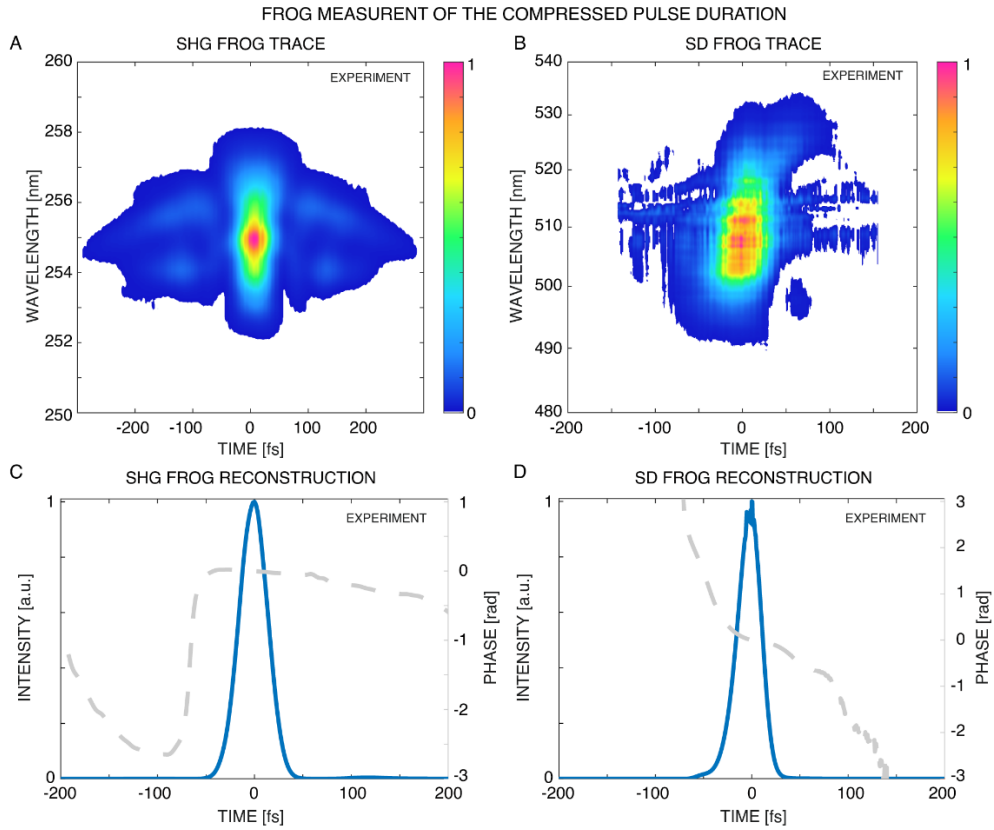

**Fig. 3. FROG measurements of the compressed sub-40 fs VIS pulse.** A and B) Experimental traces from second harmonic and self-diffraction FROGs. C and D) Corresponding retrieved 515 nm pulse intensity in the time domain. The pulse shapes show negligible femtosecond-to-picosecond pedestals.

In our experiments, the compressed pulse durations are measured using both self-diffraction FROG and second harmonic generation FROG showing some uncompensated phase stretching the pulse beyond the transform limit of 35 fs. Such multicycle pulses laser pulses (>20 cycle) in the VIS range are ideal for high harmonic generation since the phase matching window increases for short-wavelength drivers compared to mid-IR laser where under full phase matching conditions this window closes to a sub-single laser cycle.

The extremely low losses in focusing and near spatio-temporal solitary mode of propagation offers a fortunate possibility for post-laser compression at high laser energy. Scaling to higher energy and shorter-wavelength VIS-to-UV ultrafast pulses would involve using a collimated beam at the right intensity using a pair of down-scoping mirrors instead of a single lens with a long focal length. Since the beam size on the plates and the periodic spacing between plates depend on the critical phase on each plate, the soliton propagation is scalable. To estimate the parameters for spectral broadening at the fundamental and its perturbative harmonic (SHG, THG, FHG) of ultrafast Ti:Sapphire and Yb-based lasers at high energy, we assume a critical spectral phase of $B = 1.4\ rad$. We choose $CaF_2$ thin plates with a nominal thickness of 500 $\mu m$ and 1000 $\mu m$ as the Kerr material and a plate spacing of 50 mm for scaling purposes [26]. The results are summarized in Table 1. In all estimates, we note that the dispersion lengths are much longer than the plate thicknesses $L_D \gg \ell$, indicating that the pulse propagation is mainly affected by nonlinearity in the positive GDD dispersion medium as the nonlinear lengths are much smaller than the plate thicknesses $L_{NL} \ll \ell$.

Table 1. Parameters for UV-VIS-IR pulse broadening using Ti:Sapphire and Yb-based lasers and their perturbative harmonics (SHG, THG, FHG).

| Wavelength [nm] | 800 | 400 | 266 | 1030 | 515 | 515 | 258 |
|---|---|---|---|---|---|---|---|
| $n_2$ $10^{-16}\ [cm^2/W]$ | 1.80 | 2.50 | 10 | 1.70 | 2.30 | 2.30 | 10 |
| Input Pulse Duration [$fs$] | 30 | 30 | 30 | 250 | 180 | 200 | 240 |
| Pulse Energy [$mJ$] | 7 | 2.4 | 1 | 10 | 0.056 | 1.6 | 1.1 |
| Beam radius on plates $\omega$ [$\mu m$] | 2425.55 | 2354.75 | 3693.67 | 1217.46 | 176.03 | 631.18 | 1391.68 |
| Plate thickness $\ell$ [$\mu m$] | 426.453 | 422.015 | 414.311 | 854.366 | 848.65 | 424.325 | 413.273 |
| $L_s$ [$mm$] | 50 | 50 | 50 | 50 | 50 | 50 | 50 |
| Laser Intensity on Thin Plates [$TW/cm2$] | 2.37 | 0.86 | 0.15 | 1.61 | 0.60 | 1.20 | 0.142 |
| $P$ [$MW$] | 219202 | 75154.98 | 31314.58 | 37577.49 | 292.2694 | 7515.498 | 4305.754 |
| $P_{cr}$ [$MW$] | 4.00 | 7.19E-1 | 7.95E-2 | 7.01 | 1.30 | 1.30 | 7.42E-2 |
| $\frac{P}{P_{cr}}$ | 5.48E4 | 1.04E5 | 3.94E5 | 5.36E3 | 2.25E2 | 5.803 | 5.80E4 |
| $\frac{\omega^2}{\lambda L_s}$ | 147.0829 | 277.2424 | 1025.802 | 28.78062 | 1.203315 | 15.47119 | 150.7211 |
| $L_D$ [$mm$] | 3.23E1 | 1.33E1 | 7.04 | 3.383 | 6.65E2 | 8.22E2 | 4.22E2 |
| $L_{NL}$ [$mm$] | 2.98E-1 | 2.95E-1 | 2.90E-1 | 5.98E-1 | 5.94E-1 | 2.97E-1 | 2.89E-1 |
| $B$ [$rad$] | 1.43 | 1.43 | 1.43 | 1.43 | 1.43E | 1.43 | 1.43 |

Finally, we perform a VIS-UV pulse propagation simulation through 8 thin plates using 515 nm and 258 nm laser pulses (Hussar) [27]. The contribution of the plates to spectral broadening after the fourth plate is decreasing as observed in the experiments, with some distortions in the time domain. Nevertheless, these simulations allow us to extract the phase, which is then decomposed into high chromatic dispersion orders (see Table 2). These values are used as the basis for the simulation and design of prism-pair compressors and custom chirp mirrors for fine compensation of the first several orders of dispersion, Fig. 2C [21-23].

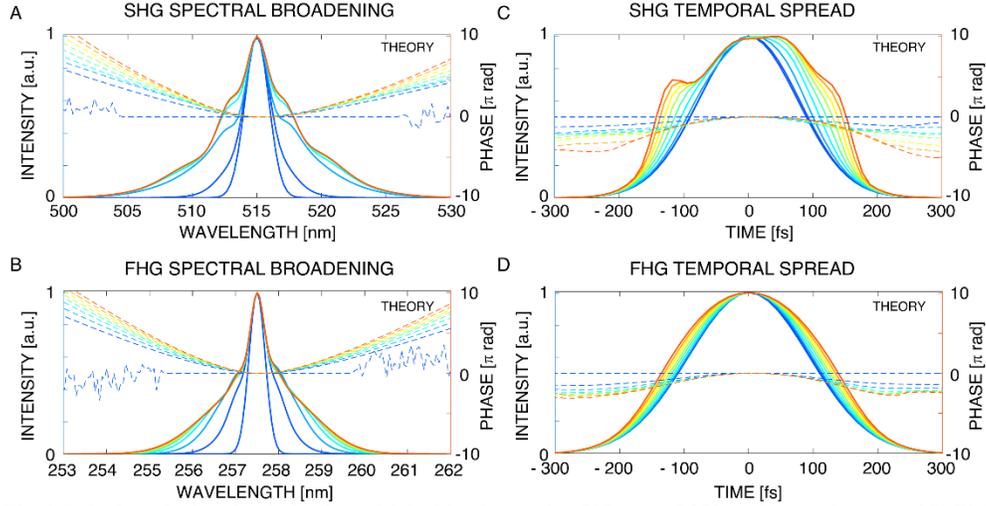

**Fig. 4. Pulse broadening simulation in multiple thin plates using 515 nm and 258 nm laser pulses.** A and B) Eight plates separated by 5 cm are illuminated by VIS-UV beams with an intensity of near ~6 $10^{11} W/cm^2$. The spectral distribution after each plate is shown in for 515 nm (A) and for 258 nm (B). C and D). The temporal spread after each plate is shown in matching colors in for 515nm (C) and for 258 nm (D).

**Table 2. Numerically extracted dispersion orders from the simulated VIS spectral broadening at 515 nm.**

| Plates | 1 | 2 | 3 | 4 | 5 | 6 | 7 | 8 |
|---|---|---|---|---|---|---|---|---|
| GDD [$fs^2$] | 1965.30 | 2177.43 | 1908.10 | 1906.33 | 1984.99 | 2076.44 | 2171.23 | 2267.30 |
| TOD [$fs^3$] | 4983.74 | 7028.24 | 3572.69 | 1377.66 | 737.89 | 607.20 | 579.98 | 590.04 |
| FOD [$fs^4$] | -21641.88 | -28921.41 | -16860.46 | -9654.81 | -7768.13 | -7585.83 | -7751.99 | -8043.90 |
| FiOD [$fs^5$] | 6.399E+04 | 8.442E+04 | 5.078E+04 | 3.108E+04 | 2.611E+04 | 2.584E+04 | 2.653E+04 | 2.756E+04 |
| SiOD [$fs^6$] | -1.816E+05 | -2.381E+05 | -1.452E+05 | -9.137E+04 | -7.806E+04 | -7.762E+04 | -7.982E+04 | -8.297E+04 |
| SeOD [$fs^7$] | 5.326E+05 | 6.961E+05 | 4.280E+05 | 2.733E+05 | 2.356E+05 | 2.349E+05 | 2.417E+05 | 2.513E+05 |
| EOD [$fs^8$] | -1.654E+06 | -2.157E+06 | -1.333E+06 | -8.595E+05 | -7.449E+05 | -7.437E+05 | -7.658E+05 | -7.964E+05 |
| NOD [$fs^9$] | 5.481E+06 | 7.138E+06 | 4.427E+06 | 2.873E+06 | 2.499E+06 | 2.497E+06 | 2.573E+06 | 2.676E+06 |
| TeOD [$fs^{10}$] | -1.941E+07 | -2.525E+07 | -1.570E+07 | -1.024E+07 | -8.929E+06 | -8.931E+06 | -9.202E+06 | -9.571E+06 |

## 4. Ultrabright High Harmonic Generation Using Short-Pulse Short-Wavelength UV-VIS Lasers

Femtosecond Yb-based lasers have experienced a surge in popularity within academic research and industry over the past decade. Their scalability in terms of high repetition rates, power output, and stability has made them particularly suitable for extreme nonlinear processes, including secondary light sources based on high harmonic generation. Recent studies have revealed that employing shorter wavelength laser drivers can significantly enhance the efficiency of high-order harmonic generation [9, 15, 16]. The single atom yield of high-order harmonics varies depending on the experimental configuration, ranging from $\lambda_{LASER}^{-5.5}$ when using constant intensity, or $\lambda_{LASER}^{(-7.5,-9)}$ for phase-matched cases where the driving laser intensity changes with the laser wavelength. By reducing the driver wavelength by a factor of 2, the single atom yield can increase by one to three orders of magnitude depending on the driving laser wavelength. Additionally, the plasma dispersion, which hampers the phase matching for long infrared drivers can be effectively compensated by leveraging the dispersion of ions for drivers in the UV-VIS spectral range. This compensation expands the temporal phase matching window for UV-VIS drivers, enabling the highly efficient generation of high-order harmonics using lasers with multiple cycles (10-to-50 pulse duration).

Notably, compressing a long 180 fs second harmonic pulse in a single self-phase modulation (SPM) stage, rather than compressing the fundamental laser beam, can yield higher second harmonic conversion efficiency while significantly improving laser quality especially in the time domain. This improvement is vital for an efficient high harmonic generation. We have performed calculations to estimate the tunnel ionization induced by the 515 nm laser with the compressed pulses using the Ammosov-Delone-Krainov (ADK) approach. Phase matching of high harmonic generation using UV-VIS lasers favors longer pulse multi-cycle durations since the temporal phase-matching window for efficient upconversion in the EUV – X-ray regime increases with the decrease of the laser wavelength. The demonstrated straightforward compression scheme provides a driver of 23-cycle pulse duration and a feasible peak intensity of $> 1.0x10^{15} \frac{W}{cm^2}$. This compression setup is expected to enables emission at $> 100\ eV$ in the EUV - soft X-ray range, at $100\ kHz$ and higher. These repetition rates are two orders of magnitude greater compared to those achieved by most conventional laser amplifiers. Theoretically, with the use of Ar ions or neutral He (as shown in Fig. 4), the emission can reach the technologically relevant $13.5\ nm$ EUV wavelength with very narrow line widths.

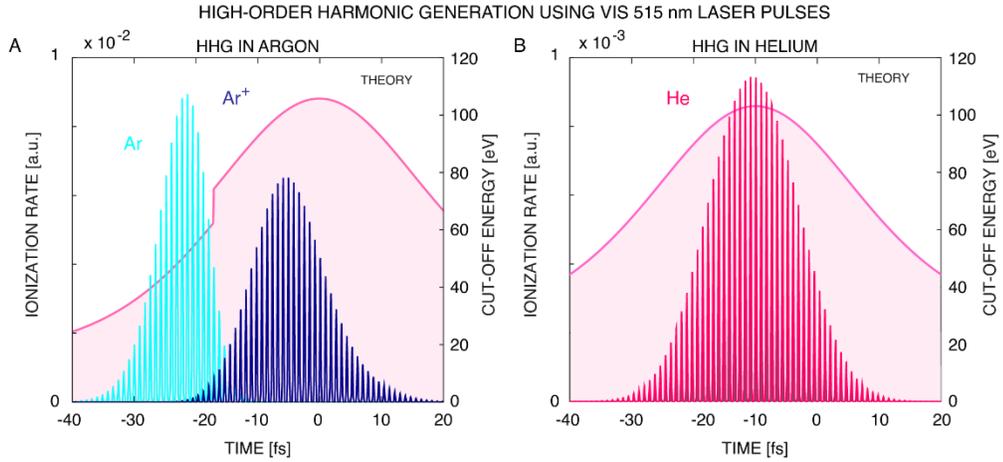

**Fig. 5. Single atom high harmonic generation at >100 eV and 100 kHz using a 23-cycle VIS driver.** Theoretical high harmonic cutoffs near the technologically relevant 13.5 nm EUV wavelength (91.7 $eV$) for a 515 $nm$ driver with a 40 $fs$ pulse duration for an experimentally feasible peak intensity of $1.0x10^{15}\ W/cm^2$ (ionization of $Ar$ atoms and ions in blue and neutral $He$ gas in red).

## 5. Conclusion

In summary, we present a robust single-stage pulse compression technique in the visible spectrum that achieves unprecedented performance in terms of spectral broadening efficiency, temporal compression efficiency, spatial beam quality, and temporal pulse quality. Our technique utilizes second-harmonic pulses at $515\ nm$ generated by a $1030\ nm$ femtosecond Yb-based regenerative amplifier and a simple multi-plate scheme with low nonlinearity that enables compression from $180\ fs$ to $40\ fs$ with minimal energy loss and distortion. Our technique offers a simple and robust solution for generating high-quality VIS pulses with durations close to the transform limit and peak powers exceeding 10 MW. Such pulses are ideal for driving extreme nonlinear optics processes and generating ultrabright EUV and X-ray high-harmonic radiation at high repetition rates of 10-100 kHz and above. Furthermore, this technique can be extended to shorter VIS and UV wavelengths and pulse durations by adjusting the plate number and thickness, opening up new possibilities for ultrafast science and technology. These features make the single-stage scheme an appealing frontend for ultrabright attosecond high harmonic generation using VIS-UV driving lasers.

**Funding.** H2020 European Research Council(100010663), XSTREAM-716950, Alfred P. Sloan Foundation(100000879)FG-2018-10892.

**Acknowledgments.** TP acknowledges funding from the European Research Council (ERC) under the European Union's Horizon 2020 research and innovation program (grant agreement XSTREAM-716950), and from the Alfred P. Sloan Foundation (FG-2018-10892).

**Disclosures.** The authors declare no conflict of interest.

**Data availability.** Data underlying the results are presented in the plotted graphs.